\documentclass[10pt,journal,comsoc]{IEEEtran}
\usepackage{algorithm}
\usepackage{algpseudocode}
\usepackage{amsmath}
\usepackage{amssymb}
\usepackage{amsthm}
\usepackage{array}
\usepackage{booktabs}
\usepackage{float}
\usepackage{graphicx}
\usepackage{subfig}
\usepackage{bm}
\usepackage{hyperref}
\usepackage{lineno}
\usepackage{listings}
\usepackage{multirow}
\pdfoutput=1
\usepackage{threeparttable}
\usepackage{url}
\usepackage{verbatim}
\usepackage{xcolor}
\usepackage{ragged2e}
\usepackage{cite}
\usepackage{diagbox}
\newtheorem{exmp}{Example}%[section]

\newtheorem{myDef}{Definition}

\ifCLASSINFOpdf
\else
\fi

\begin{document}
\title{Spatial-Aware Local Community Detection Guided by Dominance Relation}   
\author{Li Ni, Hefei Xu, Yiwen Zhang and Wenjian Luo       
        
\thanks {This work was supported by the National Key Research and Development Plan of China [No.2019YFB1704101], and the National Natural Science Foundation of China [No.62106004, No.U1936220 and No.61872002] (\textit{Corresponding author: Yiwen Zhang.})}

\thanks {Li~Ni, Hefei Xu, and Yiwen Zhang are with the School of Computer Science and Technology, Anhui University, Hefei, Anhui 230601, China. (e-mail: nlcs@mail.ustc.edu.cn, e20301226@stu.ahu.edu.cn, zhangyiwen@ahu.edu.cn).}

\thanks {Wenjian~Luo is with School of Computer Science and Technology, Harbin Institute of Technology, Shenzhen 518055, China. (e-mail: luowenjian@hit.edu.cn).}

\thanks {Note: This work has been submitted to the IEEE for possible publication. Copyright may be transferred without notice, after which this version may no longer be accessible.}
}

\maketitle
\begin{abstract}
	\justifying
The problem of finding the spatial-aware community for a given node has been defined and investigated in geo-social networks. However, existing studies suffer from two limitations: a) the criteria of defining communities are determined by parameters, which are difficult to set; b) algorithms may require global information and are not suitable for situations where the network is incomplete. Therefore, we propose spatial-aware local community detection (SLCD), which finds the spatial-aware local community with only local information and defines the community based on the difference in the sparseness of edges inside and outside the community. Specifically, to address the SLCD problem, we design a novel spatial aware local community detection algorithm based on dominance relation, but this algorithm incurs high cost. To further improve the efficiency, we propose an approximate algorithm. Experimental results demonstrate that the proposed approximate algorithm outperforms the comparison algorithms.
\justifying
\end{abstract}


\section{Introduction} \label{sec:introduction}
With the increasing popularity of location-based services, geosocial networks have emerged \cite{wang2018efficient}. Geosocial networks contain users' social relations and geographic location information. In geosocial networks, one of the most critical tasks is detecting spatial-aware communities \cite{zhu2017geo}, which has broad application prospects in many location-based social services, such as event recommendation, social marketing, and geosocial data analysis \cite{fang2017effective}.

In this paper, we study the problem of spatial-aware local community detection (SLCD) in geosocial networks. Specifically, for a geosocial network and a given node,
the objective is to find the spatial-aware local community to which the given node belongs.
The following two properties hold: 1) only local information is used in the process of detecting the community; and 2) the community satisfies both structural and spatial cohesiveness. Structural cohesiveness means that nodes inside the community are relatively tightly connected,  and nodes inside and outside the community are relatively sparsely connected, while spatial cohesiveness means that the locations of nodes in the same community are close to each other.

\textbf{Prior Work.} The studies on finding communities contain global community detection \cite{newman2004fast,2009Community}, local community detection \cite{ChenLZLYW18,LuoZNL21} and community search \cite{sozio2010community,YaoC21}.
Global community detection algorithms aim to detect all communities in social networks \cite{newman2004fast,LiHWLH20,NathSV21}.
Most global community detection studies only use topology information to detect communities \cite{2009Community, guimera2005functional}.
In real life, users' spatial location can affect social relationships because offline social activities are constrained by geography \cite{guo2008regionalization, expert2011uncovering, chen2015finding}.
Therefore, some work has considered the user's location information \cite{expert2011uncovering, chen2015finding, wang2019research,zhang2016engagement}.
Global community detection methods often require global information of the network, such as the total number of edges \cite{LuoLNZD20}. However, because of trade secrets, global information about entire networks may be unavailable or expensive to obtain. In addition, when users want to know the local community to which the given node belongs, it is not necessary to mine all the communities in the network \cite{LuoLNZD20}.

To compensate for these shortcomings, local community detection has been investigated, which can quickly detect the community that contains the given node with only local information \cite{clauset2005finding,luo2008exploring,LuoZNL21,LuoBYLHZ20}.
Similar to the local community detection problem, community search aims to find a subgraph containing a set of given nodes. Some community search works need global information of social networks \cite{sozio2010community}, and some do not \cite{cui2014local,viswanath2010analysis}. These above studies consider only the link between nodes \cite{clauset2005finding, luo2008exploring, sozio2010community, LuoZNL21,cui2014local}, ignoring the nodes' locations, so the detected communities may not be spatially cohesive and may not be suitable for some location-based services \cite{barthelemy2011spatial,chen2015finding,fang2017effective}.

To obtain a community that satisfies both structural and spatial cohesiveness, spatial-aware community search has attracted attention \cite{fang2017effective,wang2018efficient}.
Existing studies adopt different spatial constraints to restrict the geographic location of nodes to ensure spatial cohesiveness and place a minimum degree constraint on nodes to guarantee structural cohesiveness \cite{fang2017effective, zhu2017geo, wang2018efficient,Luo0XQXJ20}.
For example, Fang et al. regarded a $k$--core structure with the minimum coverage circle (MCC) as a spatial-aware community \cite{fang2019spatial}, which considers the location information of nodes and could obtain a spatially cohesive community.

\begin{figure}[!t]
\label{LBSN}
	\centering
	\includegraphics[width=0.85\linewidth]{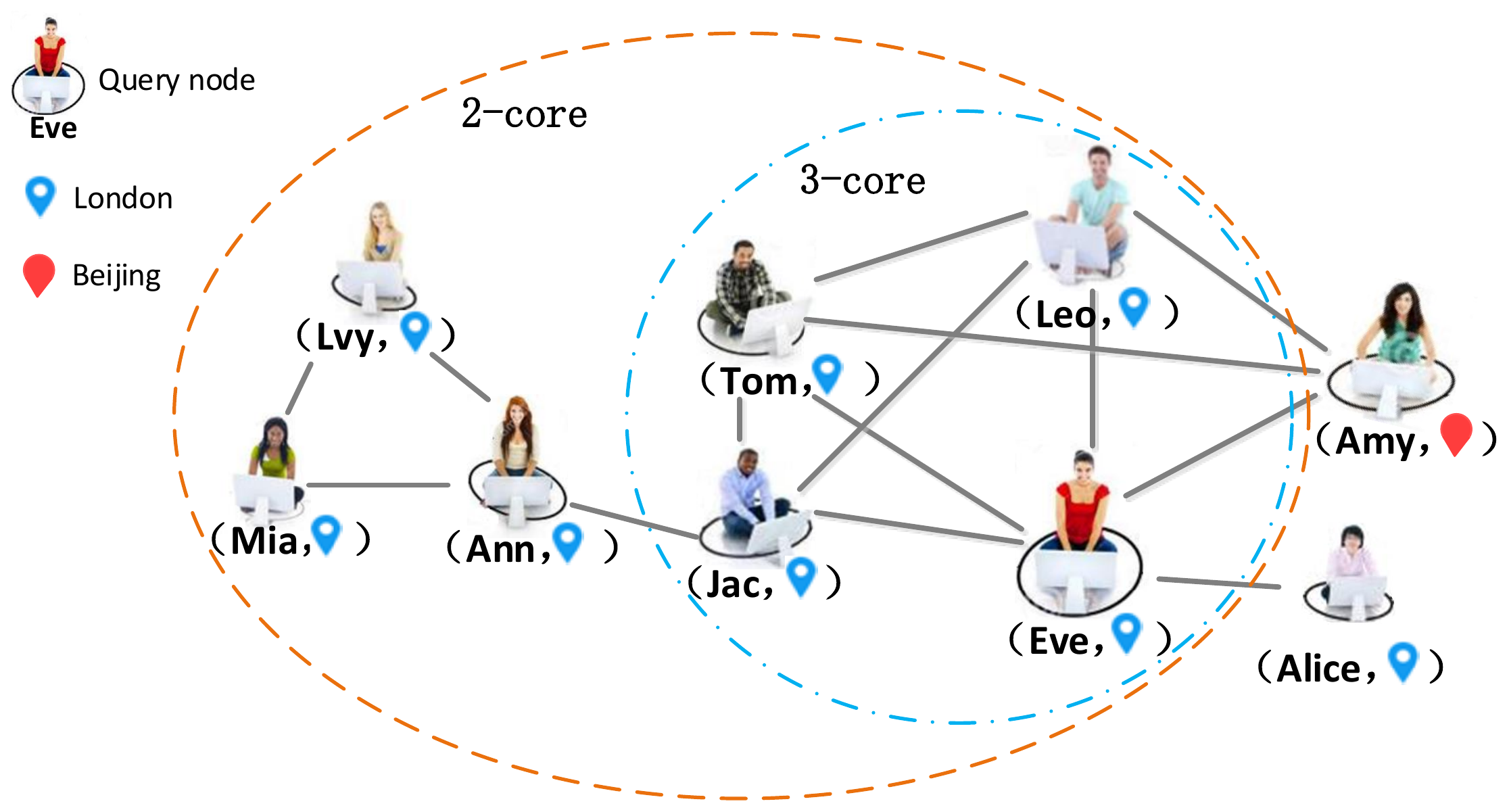}
\caption{Toy geosocial network and $k$--core structure detected by the $Exact$ algorithm \cite{fang2017effective}}
	\label{fig:network}
\end{figure}

However, existing studies need to set parameters such as parameter $k$ in the $k$--core, which is not easy for users to set \cite{LiuZZHXG20}.
If $k$ is set to be large, the $k$--core structure does not exist. On the other hand, if $k$ is set to be small, the $k$--core structure may not be structurally cohesive.
Taking Eve in Fig.~ \ref{fig:network} as an example, when $k\ge4$, the community that contains Eve does not exist; when $k=2$ or 1, the $k$--core structure is not sufficiently cohesive. When $k=3$, the 3-core containing Eve is suitable, but the 3-core containing Ann does not exist. In addition, some methods require global information \cite{fang2019spatial} and are not suitable for incomplete networks.
For situations in which global information about the network is unavailable or the user is only interested in the community to which the given node belongs, we propose SLCD, which uses only local information to detect the community.
To avoid setting the parameter, we adopt the difference between inside and outside the community to detect communities, and propose a parameter-free Spatial-aware Local community detection  method based on Dominance Relation (SLDR).
The SLDR algorithm can obtain the community of spatial cohesiveness and structural cohesiveness.
However, obtaining the derived communities by the community derivation process incurs a high cost.
To address this bottleneck, we propose an approximation community derivation process. The basic idea is to reduce the number of derived communities.

The contributions of this work are summarized as follows:

\begin{itemize}
\item We propose the SLCD problem and define dominance relation between communities based on structural and spatial cohesiveness of the communities.
\item We propose the SLDR algorithm without parameters, which iteratively performs community derivation and community filtration. We further propose the approximate SLDR algorithm with the approximate community derivation process.
\item We evaluate the effectiveness of the proposed algorithm on synthetic and real datasets. The experimental results show that the approximation algorithm outperforms other comparison algorithms.
\end{itemize}

The rest of the paper is organized as follows. Section \ref{section:Related Work} reviews the related works. Section \ref{section:Preliminaries} presents the preliminaries, including the local modularity and dominance relation. Section \ref{sec:algorithm} first introduces the SLCD problem and then designs the SLDR algorithm and its approximation algorithm. Section \ref{sec:Experiments} conducts experiments on synthetic and real geosocial networks. Section \ref{sec:conclusion} concludes our work.

\section{Related Work} \label{section:Related Work}
Studies on finding communities contain community detection \cite{newman2004fast} \cite{LuoZNL21} and community search \cite{cui2014local}. Community detection can be divided into global community detection and local community detection \cite{he2019krylov}.

\subsection{Global Community Detection}
Community detection aims to detect all communities in social networks \cite{2009Community,ZhuCZ20,Pizzuti18}.
Most community detection works consider only link information when detecting the community \cite{OkudaSSK21,LuoLSLZ21}.
Newman et al. \cite{newman2004fast} defined modularity and designed the greedy algorithm based on modularity to partition communities in the network.
Pizzuti \cite{Pizzuti12} designed a multi-objective evolutionary algorithm for community detection. It divides the nodes into different communities by optimizing the tightness of intral-connections inside each community and the sparsity of inter-connections between different communities.
Li et al. \cite{LiLZZZ21} propose a method based on network representation learning which combines node embedding and community embedding to detect communities in social networks.

Some works on geosocial networks consider the effect of users' spatial information, aiming at finding communities that satisfy spatial and structural cohesiveness \cite{chen2015finding}, such as the reports by \cite{wang2019research,zhang2016engagement}.
The works in \cite{chen2015finding} \cite{wang2019research} used the location information of the node to weight the edges to transform the unweighted network into a weighted network and then detect the communities.
Zhang et al \cite{zhang2016engagement} adopted $k$--core and the pairwise similarity between users based on attribute values (e.g., users' geo-locations) to guarantee the cohesiveness of a community from both structural and vertex attributes.
All of these works detect all communities from a global view rather than a local view, which is different from our work.
\subsection{Local Community Detection and Community Search}
The goal of local community detection is to obtain the
local community that contains the given node with only local information \cite{clauset2005finding,luo2008exploring}. Researchers have proposed various approaches for local community detection \cite{LiHKBH18,NiLZH20,LuoLNZD20,LyuSS21,ChengWZY21}.
Luo et al. \cite{LuoZNL21} proposed a local modularity $LQ$ and designed modularity optimization algorithms based on $LQ$.
He et al. \cite{he2019krylov} developed a community detection method based on the local spectral subspace, which is defined based on the Krylov subspace.
Lyu et al. \cite{LyuSS21} proposed an EA-based method for local community detection with some effective strategies in terms of individual representation, fitness evaluation, the local search operation, etc.
Most studies on local community detection do not consider the location information of nodes and are not suitable for detecting communities in geosocial networks.

Another related line of work on finding the community that contains the query node is community search. For a graph and a set of query nodes, the task of community search is finding a connected subgraph based on query parameters \cite{LiuZZHXG20,Wang00ZQZ21}.
Huang et al. \cite{huang2014querying} proposed a $k$-truss community model and designed a tree index to detect local communities in the network efficiently.
Liu et al. \cite{LiuZZ0Z21} studied the SCkT search problem, which aims to find a triangle-connected $k$--truss
containing query nodes with sizes no larger than a given threshold as a community. They investigated $expansion$ and $shrinking$ strategies to detect communities in networks. These studies on community search employ only link information to detect communities without considering the location information of nodes, which is not suitable for detecting communities that satisfy both structural and spatial cohesiveness.

Local community detection is similar to community search; both detect the community that contains the given/query node. The two are not identical, however. The difference between the two includes the definition of the community and whether global information is used when detecting the community.
\subsection{Spatial-aware Community Search}
Most existing studies employ the $k$-core model \cite{fang2017effective,zhu2017geo,wang2018efficient,fang2019spatial,kim2020densely} to ensure the structural cohesiveness of communities. These studies use different spatial constraints, such as the minimum MCC constraint, range constraint, and $k$-nearest neighbor constraint, to ensure the spatial cohesiveness of communities \cite{zhu2017geo,fang2019spatial}.
For example, Fang et al. \cite{fang2017effective,fang2019spatial} proposed the exact solution to detect the $k$-core structure covered by the smallest MCC as a community.
Wang et al. \cite{wang2018efficient} ensured that the nodes in the same community are geographically close by a radius-bounded circle.
However, the value of $k$ in these methods \cite{wang2018efficient,fang2019spatial,kim2020densely} is hard to set. When $k$ is set larger than the degree of the query node, the community does not exist. Even if $k$ is less than the degree of the query node, the community may not be found because the query node and its neighbors may not satisfy the minimum degree constraint; on the other hand, when $k$ is set to be small, too many nodes satisfy the minimum degree constraint, thus making the detected community possibly not structurally cohesive enough.

To our knowledge, existing studies on finding spatial-aware communities containing given nodes focus mainly on spatial-aware community search (SAC). SLCD has not received much attention. The goal of SLCD is to detect the local community that satisfies both structural and spatial cohesiveness only with local information. The differences between SAC and SLCD are evident in the following two aspects.
\begin{itemize}
\item For SAC, the criteria of the definition of structural cohesiveness are based on query parameters (e.g., $k$--core \cite{cui2014local}). In contrast, the criteria of defining structural cohesiveness for SLCD usually take advantage of the difference in the sparseness of edges inside and outside the community (e.g., local modularity).
\item  The SLCD algorithms use only local information when detecting communities, while SAC algorithms have no restriction on whether to use  global information. That is, some algorithms require global information \cite{huang2014querying}, and some do not \cite{CuiXWLW13}.
\end{itemize}
\section{Preliminaries} \label{section:Preliminaries}
In this section, we first introduce the local modularity and then introduce some relevant definitions about dominance relation.
\subsection{Local Modularity}
Luo et al. \cite{luo2008exploring} proposed the local modularity called $M$ to evaluate the quality of the community. Local modularity $M$ is based on internal and external edges of the community, defined as follows.
\begin{equation}\label{M}
	M=\frac{e_I}{e_O},
\end{equation}
where $e_I$ is the number of internal edges of the community and $e_O$ is the number of external edges of the community. If the number of internal edges is larger and the number of external edges is smaller, the quality of the community is better.

\subsection{Dominance Relation}\label{pareto}
Given the objective function space $F$ and the solution space $X$, some relevant definitions about dominance relation are as follows.
\begin{myDef}{(Dominance relation \cite{GONG20124050}).}
Given maximization objective functions: $f_1(x), f_2(x),$ $ f_3(x), \ldots, f_n(x) \in F$, two solutions: $x_1$, $x_2 \in X$, $\forall i\in \{1,2,\dots,n\},\exists j \in \{1,2,\dots,n\}$, if $f_i(x_1)\leq f_i(x_2 )$ and $f_j(x_1)<f_j(x_2)$, then solution $x_2$ dominates solution $x_1$ or solution $x_1$ is dominated by solution $x_2$, denoted as $x_1\prec x_2$.
\end{myDef}

\begin{myDef}{(Nondominated solution and dominated solution \cite{GONG20124050}).} Among solution space $X$, solution $x \in X$ is a nondominated solution or Pareto-optimal solution if it cannot be dominated by any solution in $X$. Otherwise, $x$ is a dominated solution.
\end{myDef}

Dominance relation is often used in multi-objective optimization to find nondominated solution \cite{FolinoP14}. To obtain a set of nondominated solutions, Liu et al. proposed the BNSA algorithm \cite{liu2011fast}. 
For the set of multiple solutions $\{x_1, x_2, x_3, \ldots, x_n\}$ and two maximization objective functions $f_1$ and $f_2$, the BNSA algorithm \cite{liu2011fast} first sorts solutions in $\{x_1, x_2, x_3, \ldots, x_n\}$ in descending order of the $f_1$ value. If the two solutions have the same value of $f_1$, the solutions are sorted in descending order of $f_2$ value.
Then, starting from the second solution, each solution $x_i$ is processed as follows. If it is dominated by the previous solution, $x_i$ is the dominated solution, which is deleted from sorted solutions.
Otherwise, $x_i$ is a nondominated solution, which is retained in sorted solutions.
The first solution is a nondominated solution because it has the maximum value of $f_1$, which implies that no other solution can dominate it. After all solutions except the first solution are processed, we can obtain  nondominated solutions. Supposing that $n$ is the size of the solution set, the time complexity of the sorting step and comparison step are $O(nlogn)$ and $O(n)$, respectively. Therefore, the time complexity of the BNSA algorithm is $O(nlogn)$ \cite{liu2011fast}.
\section{Proposed Algorithm}\label{sec:algorithm}
\subsection{Problem Statement and Community Dominance Relation} \label{sec:problem}
We start this subsection with an introduction to geosocial networks. Then we introduce the SLCD problem and community dominance relation.

A geosocial network is a graph with node location information. Let $G(V, E)$ represent a geosocial network, where $E$ represents the edge set and $V$ represents the node set. Each node in $V$ has location information, which is often represented by horizontal and vertical coordinates.

\textbf{Problem 1. (SLCD Problem)}
For a geosocial network and a given node, SLCD aims to find the spatial-aware local community, satisfying the following properties:

\begin{itemize}
\item \textbf{Connectivity.} The community containing the given node and nodes in the community are directly or indirectly connected.
\item \textbf{Structural cohesiveness.} The nodes inside the community are relatively tightly connected to each other, while nodes inside and outside the community are relatively sparsely connected.
\item \textbf{Spatial cohesiveness.} The locations of nodes in the same community are close to each other.
\item \textbf{Only local information.} Only local information is used when detecting the community.
For example, only nodes and edges in or near the community are accessed.
\end{itemize}

The difference between SLCD and SAC problems \cite{fang2017effective} is mainly in two aspects: 1) One is structural cohesiveness.
The criteria of structural cohesiveness of the SLCD problem are based on the difference in the sparseness of edges inside and outside the community. The criteria of structural cohesiveness of the SAC problem are based on query parameters (e.g., $k$--core). 
2) The other is that the SLCD problem uses only local information, while the SAC problem has no restrictions on global information of the network.
For better understanding, here, we take the $Exact$ algorithm \cite{fang2017effective} and the proposed method (Section \ref{sub:BasicAlgorithm}) as examples. The first step of the $Exact$ algorithm is to traverse all the nodes in the network to extract the $k$-core subgraph.
All nodes in geo-networks are accessed, so the $Exact$ algorithm uses global information. Our proposed method only accesses the nodes in or near the detected community. That is, only local information is utilized.

Here, we model the SLCD problem with two objective. The first objective is to maximize the structural cohesiveness of the community. The other is to maximize the spatial cohesiveness of the community. 
Specifically, we use local community $M$ (calculated by (\ref{M})) to measure structural cohesiveness, while $S$ is adopted for measuring spatial cohesiveness, formulated as
\begin{equation}
\left\{ 
\begin{array}{l}  
	max \; f_1=M=\frac{e_I}{e_O}  \vspace{1.3ex}   \\ 
    max \; f_2=S=-\frac{\sum_{{i, j}\epsilon {C}} d(i, j)}{|C|*(|C|-1)}, \\
\end{array} 
\right.
\end{equation}
where $|C|$ denotes the size of community $C$ and $d(i, j)$ denotes the distance between nodes $i$ and $j$. 
$S$ is a variant of average distance between nodes within the community, which is used to measure the degree of community spatial cohesiveness in \cite{fang2017effective,fang2019spatial}.  Here, we use $S$ as the optimization objective to improve the community spatial cohesiveness.
Maximizing $M$ and $S$ of the community could make the links within a community dense while the locations of nodes in the same community are close.

When detecting communities in geosocial networks, maximizing  the first objective may make the second objective worse, and vice versa. Specifically, maximizing  the first objective adds nodes that are more structurally connected to the community into the community. If the node is far away from the community, the second objective will worsen.
For example, for the community \{Led, Tom, Jac, Eve\} in Fig. \ref{fig:network} (or the community that contains the two closest nodes in geosocial networks), maximizing  the first objective adds Eve (one node) to the community, making the second objective worse.
Maximizing the second objective adds some nodes that are close to the community to the community. If these nodes are structurally sparsely connected to the community, the first objective will worsen.
In summary, the first objective and the second objective are potentially conflicting.

On the basis of the $M$ value and $S$ value of the community, we define the dominance relation between communities, nondominated community, dominated community, and derived community. 
\begin{myDef}{(Community dominance relation).}
Given community $C_1$ and community $C_2$, we say that community $C_2$ is dominated by community $C_1$ or $C_1$ dominates $C_2$ (denoted as $C_2\prec C_1$) if $M_{C_1}\ge M_{C_2}$ and $S_{C_1} > S_{C_2}$, or $S_{C_1}\ge S_{C_2}$ and $M_{C_1} > M_{C_2}$,  where $M_{C}$ ($S_{C}$) is $M$ ($S$) of the community $C$.
\end{myDef}
\begin{myDef}{(Nondominated community and dominated community).} Among a given set of communities, community $C$ is a nondominated community if it cannot be dominated by any communities. Otherwise, $C$ is a dominated community.
\end{myDef}

\begin{myDef}{(Derived community).} Given a nondominated community $C$ and its neighbor nodes set $N_C$, the community (e.g., $C \cup \{ v\}$) expanded by adding one node $v$ in $N_C$ to community $C$ is called the derived community.
\end{myDef}

\subsection{Basic Algorithm} \label{sub:BasicAlgorithm}
To address the SLCD problem, we design a novel Spatial-aware Local community detection algorithm with Dominance Relation (SLDR). We first introduce the SLDR algorithm and then provide a detailed description of its two key processes.

Table \ref{tab:commands} show some notations used in this paper.
\begin{table}[!t]
	\renewcommand\arraystretch{1.2}
	\caption{Meanings of some notations}
	\label{tab:commands}
	\begin{center}
	\small
 \begin{tabular}{p{0.14\columnwidth}<{\centering} p{0.65\columnwidth}}
 \toprule
 Notation & Meaning\\ % Table header row
 \midrule
 $G$ & geo-social network \\
 
	$C$ & local community $C$ \\
	
	$N_i$ & set of neighbor nodes of node $i$  \\
	
	$N_{C}$ & set of neighbor nodes community $C$ \\
	
	$M_{C}$ & $M$ of community $C$ \\
	
	$S_{C}$ & $S$ of community $C$ \\
	
	$ND$ & set of nondominated communities \\
	
	$NDE$&\multicolumn{1}{m{7cm}}{set of nondominated communities that have not been expanded}\\ 
	
	$HND$&\multicolumn{1}{m{7cm}}{set of nondominated communities that have been expanded} \\
	
	$D$ &\multicolumn{1}{m{7cm}}{set of derived communities that are expanded from nondominated communities}\\ 
 \bottomrule
 \end{tabular}
 \end{center}
\end{table}

\subsubsection{SLDR Algorithm}
Based on the definitions described in section \ref{sec:problem}, we propose the SLDR algorithm for the SLCD problem, as shown in Algorithm \ref{alg:SLDR}.
The basic idea of this algorithm is to maximize the $S$ and $M$ of the community by iteratively performing community derivation (section \ref{derivate}) and community filtration (section \ref{filt}). The community derivation process aims to expand the community by generating communities derived from nondominated communities. The community filtration process removes the dominated communities in derived communities to obtain the nondominated communities.

\begin{algorithm}[!t]
\caption{SLDR}
	\label{alg:SLDR}
\hspace*{0in} {\bf Input:}
$G, v$\\
\hspace*{0in} {\bf Output:}
community \ that\ $v$\ belongs to
\begin{algorithmic}[1]
\State $C \gets \{v\} $  \label{SLDR_ini1}
\State $N_C \gets N_{v}$
\State $M_C \gets 0,S_C\gets -\infty$
\State $HND \gets \emptyset$
\State $ND \gets \{C\}$
\State $NDE\gets \{C\}$ \label{SLDR_ini2}
\While{$NDE \neq \emptyset$}
\State $D\gets$ Community Derivation ($NDE$) (Alg. 2) \label{step:derivation}
\State $ND\gets$ Community Filtration ($D\cup ND$) (Alg. 3) \label{step:filtration}
\State $HND\gets (ND\cap HND)\cup ( NDE\cap ND)$   \label{step:updateHND}
\State $NDE\gets ND-HND$            \label{step:updateNDE}
\EndWhile
\State Select one community $C$ from $ND$ \label{step:select}
\end{algorithmic}
\end{algorithm}

The general process of the SLDR algorithm is as follows.
Initially, community $C$ contains $v$ and $NDE$ contains community $C$ (lines \ref{SLDR_ini1}--\ref{SLDR_ini2}).
First, Algorithm \ref{alg:SLDR} expands the communities in $NDE$ to obtain derived communities $D$ by the community derivation process (section \ref{derivate}).
Some nondominated communities in $ND$ may be dominated by communities in $D$ and become dominated communities.
Therefore, the algorithm removes the dominated communities from $D\cup ND$ to obtain nondominated communities $ND$ by the community filtration process (section \ref{filt}).
Then, the algorithm updates $HND$ (line \ref{step:updateHND}), including removing the dominated community in $HND$ (e.g., $ND\cap HND$) and adding the newly expanded nondominated community (e.g., $NDE\cap ND$) to $HND$.
Next, the algorithm obtains the nondominated community set $NDE$ by deleting the processed nondominated communities from $ND$ (line \ref{step:updateNDE}).
If $NDE$ is not empty, the algorithm continues the process of derivation and filtration. By iteratively performing community derivation and community filtration, the local communities are continuously expanded and optimized. Otherwise, the algorithm jumps out of the loop because the communities will not be optimized further at this point.
Finally, we select one community from $ND$ as the final community (line \ref{step:select}).
$ND$ contains multiple communities, which  have been sorted in descending order by the value of $M$. 
In order that the $M$ and $S$ values of the selected community are balanced, the  community  in  the  middle of $ND$ is selected. That is, the $\lceil |ND|/2\rceil$th community in $ND$ is selected where $|ND|$ is the number of communities in $ND$.

\begin{figure}[!t]
	\centering
	\includegraphics[width=4.5cm, height=4.5cm]{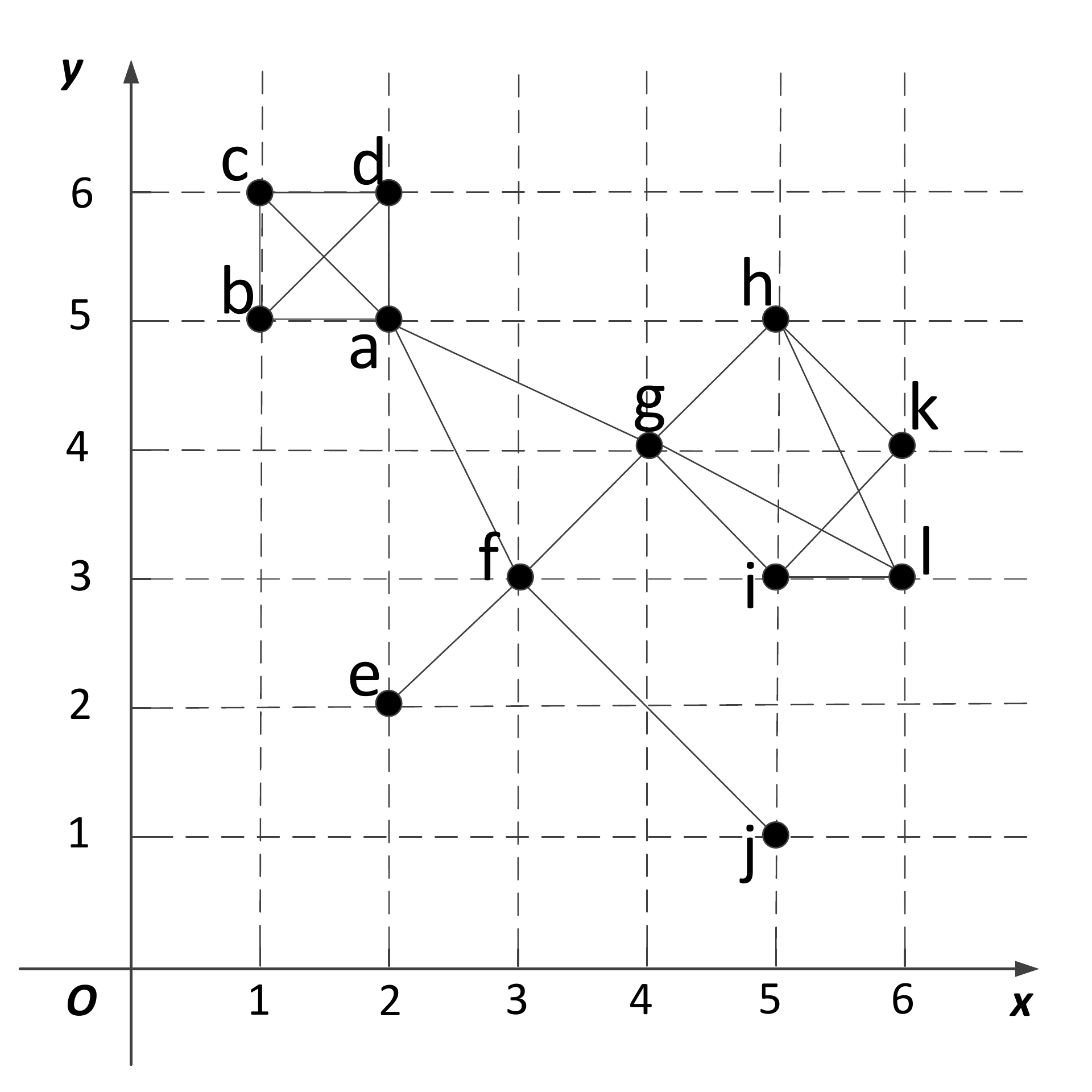}
\caption{Example of a geosocial network.
Each node has location (x, y) in a two-dimensional space	}
	\label{fig:example}
\end{figure}

\begin{figure} [!t]
	\centering
	\includegraphics[width=0.96\linewidth]{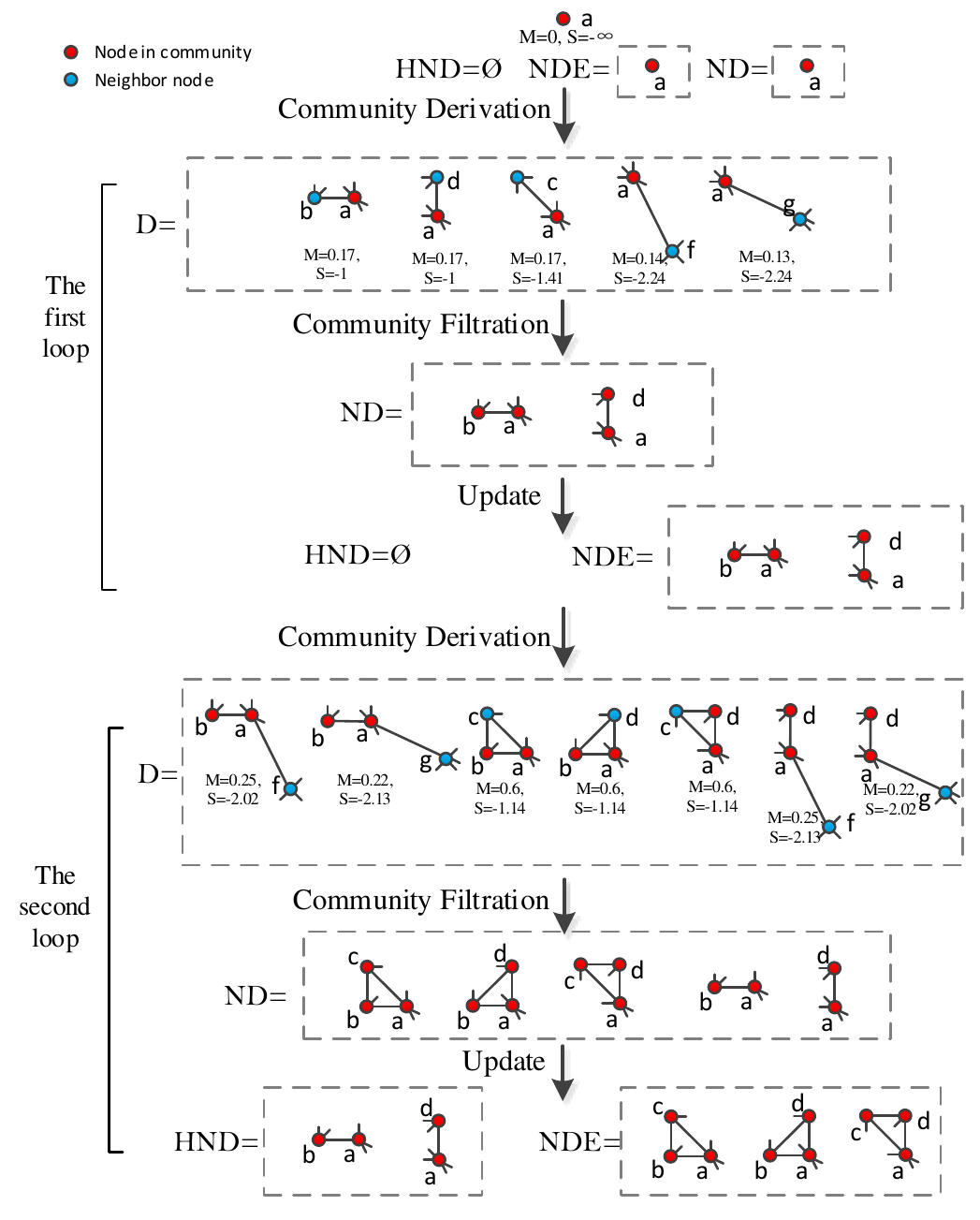}
\caption{Illustration of the SLDR algorithm}
	\label{fig:SLDR}
\end{figure}

\begin{exmp} \label{exmp:SLDR}
We use node a as the given node and the geosocial network in Fig.~ \ref{fig:example} to illustrate the process of the SLDR algorithm.
The states of $D$, $ND$, $HND$ and $NDE$ after the first and second loops are shown in Fig.~ \ref{fig:SLDR}.
Initially, the community is \{a\}, and $NDE$ contains \{a\}.
Since $NDE$ is not empty, the algorithm enters the first loop.
After derived community set $D$ is obtained,
the communities in $D\cup HND$ are screened to obtain nondominated communities $ND$ = \{\{a, b\}, \{a, d\}\}.
Then, $HND$ is empty, and $NDE\cap ND$ = \{\{a\}\} $\cap$ \{\{a, b\}, \{a, d\}\} = $\emptyset$, so $HND$ is empty.
Since community set $NDE$ = $ND-HND$ = \{\{a, b\}, \{a, d\}\} is not empty, the algorithm enters the next loop.
The second loop performs community derivation and community filtration to obtain $D$ and $ND$, respectively.
Since $(ND\cap HND)$ is empty and $NDE \cap ND$ = \{\{a, b\}, \{a, d\}\}, $HND$ is \{\{a, b\}, \{a, d\}\}. Communities in $HND$ has been processed before
and do not be added to the $NDE$.
The community set $NDE$ = \{\{a, b, c\}, \{a, b, d\}, \{a, c, d\}\}.
Since $NDE$ is not empty, the algorithm continues community derivation and community filtration processes, omitted to save space.
\end{exmp}

Although the geosocial network $G$ is the input of the SLDR algorithm, the SLDR algorithm visits only the neighbor nodes of the communities to obtain derived communities. Therefore, the proposed algorithm only uses the local information instead of the global information of the geosocial network.
\subsubsection{Community Derivation Process}\label{derivate}
The community derivation process aims to expand the community by generating derived communities of nondominated communities in $NDE$. The essential operation of obtaining a derived community is to add one neighbor node in $N_C$ to community $C$ to form a newly derived community $C'$.

\begin{algorithm}[!t]
\caption{Community\ Derivation}
	\label{alg:Derivation}
\hspace*{0in} {\bf Input:}
$NDE$\\
\hspace*{0in} {\bf Output:}
$D$
\begin{algorithmic}[1]
\State $D\gets \emptyset$  \label{step:Dinitilize}
\For{each\ community\ $C$\ in\ $NDE$}
\For{each\ node\ $u$\ in\ $N_{C}$}
\State ${C}'={C}\cup\{u\}$
\If{${C}'$\ is\ in\ $D$}
\State continue
\Else
\State compute\  $M_{{C}'}$\ and\ $S_{{C}'}$
\State $N_{{C}'}=N_C\cup N_u-C'$
\State add\ ${C}'$\ to\ $D$
\EndIf
\EndFor
\EndFor
\State \Return $D$
\end{algorithmic}
\end{algorithm}

Algorithm \ref{alg:Derivation} shows the process of community derivation, which is described as follows.
First, derived community set $D$ is empty (line 1).
Then, for each community $C$ in $NDE$ and for each node $u$ in $N_{C}$, the following steps are performed (lines 4-10):
a) The algorithm obtains the derived community $C'$ by adding node $u$ to $C$;
b) If $C'$ is not in $D$, the algorithm calculates $M_{C'}$ and $S_{C'}$, obtains $N_{C'}$, and adds the derived community $C'$ to $D$.
Finally, the derived community set $D$ is obtained.

\begin{exmp} \label{exmp:derivation}
We continue with example \ref{exmp:SLDR}.
In the first loop, initially, $NDE$ = \{\{a\}\} and $D$ = $\emptyset$.
For node b in $N_{\{{a}\}}$ where $N_{\{{a}\}}$ = \{b, c, d, f, g\}, the following steps are performed:
i) The algorithm obtains a derived community $C'$ = \{a, b\};
ii) Since \{a, b\} is not in $D$,
the algorithm computes $M_{\{a, b\}}$ and $S_{\{a, b\}}$ to be 0.17 and -1, respectively;
iii) The algorithm obtains $N_{\{a, b\}}$ = \{c, d, f, g\} and adds the derived community \{a, b\} to $D$.
Similarly, c, d, f and g are added to community $C$. Therefore, $D$ = \{\{a, b\}, \{a, d\}, \{a, c\}, \{a, f\}, \{a, g\}\}.
\end{exmp}
\subsubsection{Community Filtration Process}\label{filt}
In the community filtering process, we apply BNSA \cite{liu2011fast} (Section \ref{pareto}) to obtain nondominated communities by removing dominated communities from derived communities.

\begin{algorithm}[!t]
\caption{Community\ Filtration}
\label{alg:Filtration}
\hspace*{0in} {\bf Input:}
$D\cup ND$\\
\hspace*{0in} {\bf Output:}
$ND$
\begin{algorithmic}[1]
\State $ND \gets$ sort communities in $D\cup ND$
\For{each\ community\ $C$\ in\ $ND$}
\If{$C\prec C^{prior}$}
\State remove\ $C$\ from\ $ND$
\Else
\State continue
\EndIf
\EndFor
\State \Return $ND$
\end{algorithmic}
\end{algorithm}
Algorithm \ref{alg:Filtration} shows the process of community filtration.
First, the algorithm sorts the communities in $D\cup HND$ in descending order of $M$ and $S$ values to obtain the sorted list $ND$ (line 1).
If the $M$ values of two communities are equal, the communities are sorted in descending order of $S$ value. For convenience, for community $C$, let $C^{prior}$ represent the previous community of $C$. Then, the algorithm removes dominated communities (lines 2-6). Specifically, starting with the second community in $ND$, each community is processed as follows: If community $C$ is dominated by $C^{prior}$, then community $C$ is a dominated community, so the algorithm removes $C$ from $ND$; otherwise, community $C$ is a nondominated community, which is retained in $ND$. The first community in $ND$ is a nondominated community because it has the largest $M$ among all communities in $ND$.
Finally, Algorithm \ref{alg:Filtration} returns the nondominated communities in $ND$.

\begin{exmp} \label{exmp:filtration}
Continue with the example \ref{exmp:derivation}. In the first loop, sort communities in $D\cup HND$ to obtain $ND$ = \{\{a, b\}, \{a, d\}, \{a, c\}, \{a, f\}, \{a, g\}\}. The first community \{a, b\} in $ND$ is a nondominated community.
For the second community \{a, d\}, since $M_{\{a, d\}}$ and $S_{\{a, d\}}$ are equal to $M_{\{a, b\}}$ and $S_{\{a, b\}}$, respectively, \{a, d\} is not dominated by \{a, b\}. Therefore, \{a, d\} is a nondominated community.
For the third community \{a, c\}, since  $S_{\{a, c\}}$ is less than $S_{\{a, d\}}$, \{a, c\} is not a nondominated community. Community \{a, c\} is removed.
Similarly, \{a, f\} and \{a, g\} are dominated communities, which are removed from $ND$.
\end{exmp}

\subsection{Approximate Algorithm}
In this section, we first analyze the SLDR algorithm. Then, we introduce an approximate community derivation process. In addition, the algorithm that uses the  approximate community derivation process is the proposed approximation algorithm, called AppSLDR algorithm.

In the community derivation process, for a nondominated community $C$, each neighbor node in $N_C$ is added to community $C$ to obtain a derived community. The number of derived communities of a community is equal to the number of neighbor nodes of this community.
For each derived community $C'$, the community filtration process needs to compute $M_{C'}$ and $S_{C'}$.
If there are too many nondominated communities or too many neighbor nodes of a nondominated community, community derivation and community filtration processes have high computational costs. Thus, we developed a more efficient approximate community derivation process.

\textbf{Observation.} We start with an important phenomenon. During the execution of the SLDR algorithm, we observe a phenomenon: in most cases, the number of derived communities is much larger than that of nondominated communities. This phenomenon shows that many of the communities obtained in the process of community derivation are dominated communities. This means that the community derivation process spends considerable time computing those communities that will be eliminated in the community filtration process.

Based on the above observation, we developed an efficient approximation community derivation process.
The basic idea is that we use part of the neighbor nodes of the nondominated community to obtain derived communities rather than all neighbor nodes. The $M$ value of the derived community, obtained by the node with few internal edges and many external edges, may be relatively small, which makes the derived community have a high probability of being a dominated community. These nodes are no longer combined with the community to generate derived communities, thereby reducing time.

The method of selecting nodes is as follows: We first sort the nodes in $N_C$ in descending order of the inward ratio \cite{he2019krylov}, which is defined as the ratio of inward edges to the out-degree.
Naturally, based on the value of inward ratio, the nodes in $N_C$ are evenly divided into upper, middle and lower levels. The nodes of the upper level usually generate higher quality communities.
So, we choose the top $\lceil |N_C|/3 \rceil$ nodes in the sorted list, termed $SN_{c}$. This method can be implemented by replacing $N_C$ with $SN_{c}$ in line 3 and replacing $N_{C'}$ with $SN_{C'}$ in line 9 of Algorithm \ref{alg:Derivation}, where $SN_{C'}$ represents the
top $\lceil |N_C|/3 \rceil$ nodes in the sorted neighbor nodes of community $C'$. 
\begin{figure}[!t]
	\centering
	\includegraphics[width=0.9\linewidth]{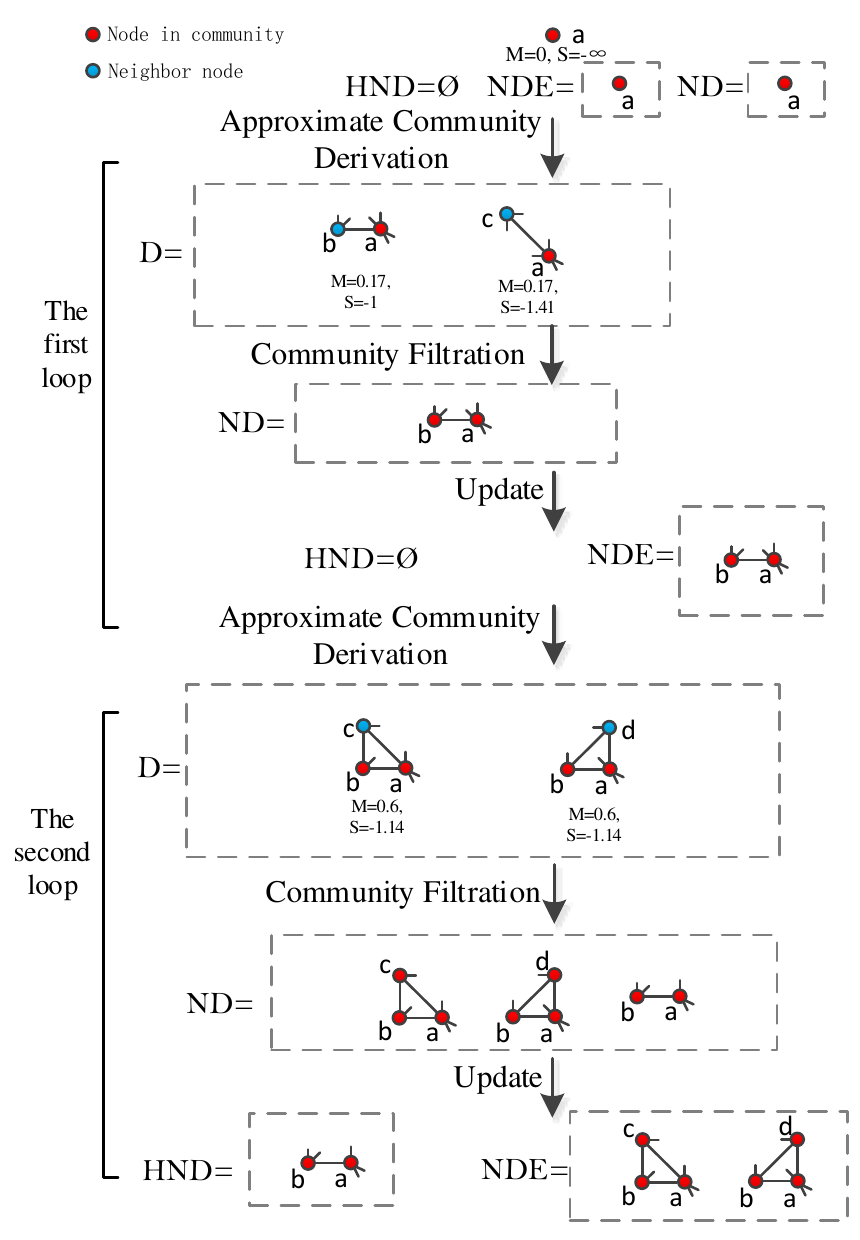}
\caption{Illustration of the AppSLDR algorithm}
	\label{fig:app}
\end{figure}

\begin{exmp}  Fig.~\ref{fig:app} shows the process of approximate community derivation.
In the first loop, Fig.~\ref{fig:app} shows two derived communities \{a, b\} and \{a, c\}, while five derived communities are generated in Fig. \ref{fig:SLDR}. Similarly, for the second loop, two derived communities are shown in Fig.~\ref{fig:app}, while seven derived communities are generated in Fig. \ref{fig:SLDR}.
Although the number of derived communities generated by the approximation process is less than that of the original community derivation,
in the second loop, $NDE$ in Fig. \ref{fig:SLDR} is similar to $NDE$ in Fig.~\ref{fig:app}, which ensures that the performance of the approximation algorithm is close to the basic algorithm.
\end{exmp}

\section{Experiments}\label{sec:Experiments}
In this section, we conduct comprehensive experiments to test the proposed algorithms. We first introduce the experimental settings, including datasets, evaluation metrics and comparison algorithms.
Implementation of this work was carried out using Centos7 (CPU: Intel(R) Xeon(R) CPU E5-2630 v3 @ 2.40 GHz, memory: 200 GB).
The algorithms were implemented using Python 3.7 programming language.
\subsection{Experimental Settings}

\subsubsection{Datasets}
We tested our algorithm on four real and two synthetic datasets. The statistics of these datasets are summarized in Table \ref{tab:datasetsInfo}. For each dataset, we evenly select 200 given nodes for the experiments. Each node is selected as the given node for spatial-aware local community detection, and then the average values of metrics are calculated for all selected nodes.
\begin{table*}[!t]
	\caption{Statistics of datasets}
	\renewcommand\arraystretch{1.1}
	\label{tab:datasetsInfo}
	\begin{center}
	\small
	\setlength{\tabcolsep}{5mm}{
    	\begin{tabular}{c<{\centering}c<{\centering}c<{\centering}c<{\centering}c<{\centering}}
            \toprule	
    		Type & Datasets & \#Vertices & \#Edges & Average Degree \\   
    		\hline
    		\multirow{4}*{Real}
    		& $Brightkite$ & \rmfamily 51406 & \rmfamily 197167 & \rmfamily 7.67 \\
    		& $Gowalla$ & \rmfamily 107092 & \rmfamily 456830 & \rmfamily 8.53 \\
    		& $Flickr$ & \rmfamily 214698 & \rmfamily 2096306 & \rmfamily 19.5 \\
    		& $Foursquare$ & \rmfamily 2127093 & \rmfamily 8460352 & \rmfamily 8.12 \\
    		\cline{1-5}
    		\multirow{2}*{Synthetic}
    		&$Syn1$	& \rmfamily 5000 & \rmfamily 20000 & \rmfamily 8 \\
    		&$Syn2$ & \rmfamily 200000 & \rmfamily 800000 & \rmfamily 8 \\
    		\bottomrule
    	\end{tabular}}
	\end{center}
\end{table*}
The four real datasets are $Brightkite$\footnote{http://snap.stanford.edu/data/index.html\label{web}}, $Gowalla$\textsuperscript{\ref{web}}, $Flickr$\footnote{https://www.flickr.com/}, and $Foursquare$\footnote{https://archive.org/details/201309\_foursquare\_dataset\_umn}.
In the above four real datasets, each node is a user, and each edge is the friendship between two users.
We implement the processing of these datasets referring to \cite{fang2019spatial}, introduced as follows:
a) $Brightkite$ contains 4,491,143 checkin records collected from 772,783 different places from April 2008 to October 2010. The user's geographic location is the place that the user checks most frequently.
b) $Gowalla$ contains 6,442,892 checkin records collected from 1280,969 places. Similar to the $Brightkite$ dataset, the place that the user checks most often is marked as the user's geographic location.
c) $Flickr$ contains locations where photos were taken. The location where a user took photos most frequently is marked as the user's geographic location.
d) $Foursquare$ contains 33,278,683 checkin records, obtained by crawling the Foursquare website. Each user in the $Foursquare$ dataset has a location of he/her hometown position, which is regarded as the user's geographic location.

We also conducted experiments on synthetic datasets. Synthetic networks are generated by a graph generator named GTGraph\footnote{ http://www.cse.psu.edu/?madduri/software/GTgraph/}, following the method in \cite{wang2018efficient,fang2019spatial}. We obtained the synthetic datasets by the following two steps:
(1) Generate a social network without node location information by the R-MAT graph generator in GTGraph. The degrees of the nodes in the generated network obey a power-law distribution, and the default parameter values of the GTGraph are adopted;
(2) Generate location information for all nodes in the social network. We randomly generate a location coordinate for each node with a location range of $[0, 1] \times [0, 1]$. This process is repeated for each node until all nodes have location coordinates.
Based on these steps, we generated two synthetic datasets: $Syn1$ and $Syn2$.

\subsubsection{Evaluation metrics} \label{metric}
When measuring the quality of the community, we take both structural and spatial cohesiveness of the community into account. The metric $communitude$ \cite{miyauchi2015network,chakraborty2017metrics} is adopted to measure the degree of structural cohesiveness of the community; $d_{avg}$ and $d_{IO}$ are adopted to measure the degree of spatial cohesiveness of the community. These three metrics are introduced as follows:

a) The metric $communitude$ \cite{miyauchi2015network,chakraborty2017metrics} measures the structural cohesiveness based on internal and external structural differences of communities, and it is calculated as follows:
\begin{equation}
\small
	communitude(C)=\frac{\frac{e[C]}{m}-(\frac{D[C]}{2m})^2}{\sqrt{(\frac{D[C]}{2m})^2(1-(\frac{D[C]}{2m})^2)}},
\end{equation}
where $m$ is the total number of edges in the network, $e[C]$ is the number of internal edges of $C$, and $D[C]$ is the sum of degrees of the nodes in $C$. A higher value of $communitude$ indicates better structural cohesiveness of the community.

b) Here, $d_{avg}$ measures the spatial proximity of nodes in the community. It is defined as the average distance between node pairs in $C$. The smaller the value of $d_{avg}$ is, the better the spatial cohesiveness of the community.

c) In addition, $d_{IO}$ measures spatial cohesiveness based on the geographical proximity of internal nodes and external nodes of the community, which is calculated by (\ref{eql:dio}).
\begin{equation}\label{eql:dio}
	d_{IO}=\frac{d_{avg}}{\sum_{i\epsilon {C}}\sum_{j\epsilon N_C} d(i, j)/({|C|*|N_C|})},
\end{equation}
where $d(i, j)$ is the distance between node $i$ and node $j$. The smaller the value of $d_{IO}$ is, the better the spatial cohesiveness of the community.
\subsubsection{Comparison algorithms}
We compare the proposed method with a local community detection method (i.e., M method \cite{luo2008exploring}), a global spatial-aware community detection method (i.e., Geomod \cite{chen2015finding}), and a spatial-aware community search method (i.e., AppAcc \cite{fang2019spatial}). We briefly describe these comparison algorithms:
\begin{itemize}
\item M method \cite{luo2008exploring}. This method starts with a given node and then finds a community with the largest local modularity $M$.
\item Geomod \cite{chen2015finding}. Geomod is a global spatial-aware community detection algorithm that detects all communities in geosocial networks. In our experiments, we select the community containing the given node from all detected communities. In addition, parameter $n$ is set to 2.

\item AppAcc \cite{fang2017effective}. The parameters of the AppAcc algorithm follow the experimental setting in \cite{fang2017effective}. Specifically, $k$ is set to 4, and $\epsilon_A$ is set to 0.5. Since the exact algorithm ($Exact+$) in \cite{fang2017effective} is slow on large data sets, we take the AppAcc algorithm (the most accurate approximation algorithm \cite{fang2017effective}) as the comparison algorithm.
\end{itemize}
Since the SLDR algorithm runs slowly, we use the AppSLDR algorithm instead of the SLDR algorithm to compare with other algorithms.
In addition, if the running time of the AppSLDR algorithm for a node is longer than two hours, we terminate it and select one community from the current communities in $ND$ as the final community.
\subsection{Results}
Considering that communities detected by AppAcc method for some nodes are empty, when comparing M, AppAcc, Geomod, and AppSLDR algorithms, the average values of metrics are calculated for nodes whose communities detected by AppAcc are not empty. Moreover, the average values of metrics calculated for all selected nodes are also given for M, Geomod, and AppSLDR algorithms.
In addition, Geomod was executed on the $Foursquare$ dataset for more than a week and still did not terminate, so its results are not given.
\subsubsection{Structural cohesiveness}
\begin{table*}[!t]
	\fontsize{10}{15}\selectfont
	\caption{Comparison of $communitude$  for nodes whose communities detected by AppAcc are not empty. "Num" means the number of nodes whose communities detected by AppAcc are not empty}
	\renewcommand\arraystretch{1.1}
	\label{com}
	\centering
	\small
	\begin{threeparttable}\setlength{\tabcolsep}{6.65mm}{
    	\begin{tabular}{c|c|c|c|c|c}
    	 \toprule
    		\diagbox{Dataset}{Method} & M & AppAcc & Geomod & AppSLDR &Num\\
    		\hline
    		$Brightkite$ & \rmfamily 0.419 & \rmfamily 0.304 & \rmfamily 0.449 & \bfseries\rmfamily 0.521&\rmfamily 77 \\
    		
    		$Gowalla$ & \rmfamily 0.439 & \rmfamily 0.294& \rmfamily 0.431 & \bfseries\rmfamily 0.525&\rmfamily 76\\
    		
    		$Flickr$ & \rmfamily 0.262 & \rmfamily 0.128 & \rmfamily 0.190& \bfseries\rmfamily 0.279&\rmfamily 143\\
    		
    		$Foursquare$ & \rmfamily 0.406 & \rmfamily 0.256 & / & \bfseries\rmfamily 0.428&\rmfamily \rmfamily 53 \\
    		
    		$Syn1$	 & \bfseries\rmfamily 0.352 & \rmfamily 0.106 &\rmfamily 0.211 & \rmfamily 0.349&\rmfamily 167 \\
    		
    		$Syn2$ & \rmfamily 0.320 & \rmfamily 0.080 & \rmfamily 0.153 & \bfseries\rmfamily 0.324&\rmfamily 181 \\
    		\bottomrule
    	\end{tabular}}
    	\begin{tablenotes}
        \footnotesize      
        \item[*] "/" means that the result is not given.
      \end{tablenotes}
	\end{threeparttable}
\end{table*}

\begin{table*}[!t]
	\fontsize{10}{15}\selectfont
	\caption{Comparison of $communitude$ for all selected nodes}
	\renewcommand\arraystretch{1.1}
	\label{com2}
	\centering
	\small
	\begin{threeparttable}\setlength{\tabcolsep}{6.65mm}{
    	\begin{tabular}{c|c|c|c}
    	 \toprule
    		\diagbox{Dataset}{Method} & M & Geomod & AppSLDR\\
    		\hline
    		$Brightkite$ & \rmfamily 0.510&\rmfamily 0.507 &\bfseries\rmfamily 0.550 \\
    		
    		$Gowalla$ & \rmfamily 0.529 &\rmfamily 0.490 &\bfseries\rmfamily 0.530\\
    		
    		$Flickr$ &  \bfseries\rmfamily 0.287&\rmfamily 0.198 &\rmfamily 0.276\\
    		
    		$Foursquare$ & \bfseries\rmfamily 0.464 &/ &\rmfamily 0.411 \\
    		
    		$Syn1$	 & \bfseries\rmfamily 0.361 &\rmfamily 0.211 & \rmfamily 0.353 \\
    		
    		$Syn2$ & \rmfamily 0.322 &\rmfamily 0.154 &\bfseries\rmfamily 0.325 \\
    		\bottomrule
    	\end{tabular}}
    	\begin{tablenotes}
        \footnotesize      
        \item[*] "/" means that the result is not given.
      \end{tablenotes}
	\end{threeparttable}
\end{table*}

Metric $communitude$ (Section \ref{metric}) is adopted to evaluate the structural cohesiveness of the community.
Table \ref{com} shows the average $communitude$ of the communities detected for nodes whose communities detected by AppAcc are not empty.
Table \ref{com2} shows the average $communitude$ of the communities detected for all selected nodes.

Table \ref{com} shows that, in most cases, the AppSLDR algorithm outperforms the AppAcc, M and Geomod algorithms.
The $communitude$ value of the AppSLDR algorithm is more than 2 times that of the AppAcc method.
The AppSLDR algorithm uses the idea of maximizing both goals of the community, which can climb out of a local optimum to find a community with better structural cohesiveness.
The M method performs better than AppAcc and Geomod, as it is designed only for the link-based analysis of nodes and does not consider the spatial cohesiveness of the community, so it focuses more on detecting communities with better structural cohesiveness.
From the table \ref{com}, we see that the values of AppAcc is relatively low. The reason is that the AppAcc algorithm guarantees only the closeness of connections within the community without considering the sparsity of the connections inside and outside the community.
The performance of Geomod is worse than that of our algorithm. This is because Geomod uses a global modularity (i.e., $Q^{geo}$ \cite{chen2015finding}) to partition the network into several communities to find
%Editor: Please ensure that the intended meaning has been maintained in this edit.
the global optimum. 

Table \ref{com2} shows the AppSLDR algorithm is competitive with  M method, and better than the Geomod method, which indicates that AppSLDR algorithm does not lose the structural cohesiveness of the community due to the the consideration of  spatial cohesiveness. 
Compared with Table \ref{com}, we notice some numerical fluctuations in the experimental results on the $Brightkite$, $Gowalla$ and $Foursquare$ dataset.
It is because many nodes with empty community detected by the AppAcc algorithm  are not considered in Table \ref{com}. 
For example,  for the $Brightkite$ dataset, Table \ref{com} shows the average $communitude$ of 77 nodes whose communities detected by AppAcc are not empty, and Table \ref{com2} shows the the average $communitude$ of all  200 selected nodes. 
On the $Syn2$ and $Flickr$ datasets, the difference between the $communitude$ values in Table \ref{com} and that in Table \ref{com2}  is small.
\subsubsection{Spatial cohesiveness}
\begin{table*}[!t] 
	\caption{Comparison of $d_{avg}$ and $d_{IO}$ for nodes whose communities detected by AppAcc are not empty}
	\renewcommand\arraystretch{1.1}
	\label{spatail}
	\centering
	\fontsize{10}{15}\selectfont
	\small
	\setlength{\tabcolsep}{4.1mm}{
	\begin{tabular}{c|c|c|c|c|c|c|c|c}
		\toprule
		\multirow{2}{*}{\diagbox{Dataset}{Method}}&
		\multicolumn{2}{c|}{M}&\multicolumn{2}{c|}{AppAcc}&\multicolumn{2}{c|}{Geomod}&\multicolumn{2}{c}{AppSLDR}\cr\cline{2-9}
		&$d_{avg}$&$d_{IO}$&$d_{avg}$&$d_{IO}$&$d_{avg}$&$d_{IO}$&$d_{avg}$&$d_{IO}$\cr
		\hline
		$Brightkite$ & \rmfamily 0.098 & \rmfamily 0.796 & \rmfamily 0.024& \rmfamily 0.128& \rmfamily 0.061&\rmfamily 0.378 & \bfseries\rmfamily 0.009 &\bfseries\rmfamily 0.070\cr
		
		$Gowalla$  &\rmfamily 0.045 &\rmfamily 0.620& \rmfamily 0.017 &\rmfamily 0.108& \rmfamily 0.037 &\rmfamily 0.295& \bfseries\rmfamily 0.005& \bfseries\rmfamily 0.074 \cr
		
		$Flickr$  & \rmfamily 0.172 &\rmfamily 0.780& \rmfamily 0.036 &\rmfamily 0.139& \rmfamily 0.105& \rmfamily 0.415& \bfseries\rmfamily 0.014 &\bfseries\rmfamily 0.091\cr
		
		$Foursquare$  & \rmfamily 0.066 &\rmfamily 0.836& \rmfamily 0.012 &\bfseries\rmfamily 0.075& / &/& \bfseries\rmfamily 0.005 &\rmfamily 0.094\cr
		
		$Syn1$	& \rmfamily 0.422 &\rmfamily 0.974& \rmfamily 0.288 &\rmfamily 0.705& \rmfamily 0.353 &\rmfamily 0.787& \bfseries\rmfamily 0.285 &\bfseries\rmfamily 0.704\cr
		
		$Syn2$  & \rmfamily 0.433 &\rmfamily 1.001& \bfseries\rmfamily 0.240 &\bfseries\rmfamily 0.612& \rmfamily 0.381 &\rmfamily 0.861& \rmfamily 0.275 &\rmfamily 0.704\cr
		\bottomrule
	\end{tabular}}
	\begin{tablenotes}
        \footnotesize      
        \item[*] "/" means that the result is not given.
      \end{tablenotes}
\end{table*}
\begin{table*}[!t] 
	\caption{Comparison of $d_{avg}$ and $d_{IO}$ for all selected nodes}
	\renewcommand\arraystretch{1.1}
	\label{spatail2}
	\centering
	\fontsize{10}{15}\selectfont
	\small
	\setlength{\tabcolsep}{6.1mm}{
	\begin{tabular}{c|c|c|c|c|c|c}
		\toprule
		\multirow{2}{*}{\diagbox{Dataset}{Method}}&
		\multicolumn{2}{c|}{M}&\multicolumn{2}{c|}{Geomod}&\multicolumn{2}{c}{AppSLDR}\cr\cline{2-7}
		&$d_{avg}$&$d_{IO}$&$d_{avg}$&$d_{IO}$&$d_{avg}$&$d_{IO}$\cr
		\hline
		$Brightkite$ & \rmfamily 0.083&\rmfamily 0.799 & \rmfamily 0.062&\rmfamily 0.386 &\bfseries\rmfamily 0.012 & \bfseries\rmfamily 0.107\cr
		
		$Gowalla$ & \rmfamily 0.045&\rmfamily 0.627 & \rmfamily 0.037 &\rmfamily 0.311&\bfseries\rmfamily 0.005 & \bfseries\rmfamily 0.105\cr
		
		$Flickr$  & \rmfamily 0.164&\rmfamily 0.743 & \rmfamily 0.103&\rmfamily 0.408 &\bfseries\rmfamily 0.014 & \bfseries\rmfamily 0.094\cr
		
		$Foursquare$  &\rmfamily 0.076 &\rmfamily 0.845 & /& /&\bfseries\rmfamily 0.021 &\bfseries\rmfamily 0.183 \cr
		
		$Syn1$	&\rmfamily 0.419&\rmfamily 0.967 &\rmfamily 0.355 &\rmfamily 0.790 &\bfseries\rmfamily 0.285 &\bfseries\rmfamily 0.702 \cr
		
		$Syn2$  &\rmfamily 0.435 &\rmfamily 1.009 & \rmfamily 0.382&\rmfamily 0.863 &\bfseries\rmfamily 0.275 & \bfseries\rmfamily 0.706\cr
		\bottomrule
	\end{tabular}}
	\begin{tablenotes}
        \footnotesize      
        \item[*] "/" means that the result is not given.
      \end{tablenotes}
\end{table*}
Here, $d_{avg}$ and $d_{IO}$ are adopted to measure the spatial cohesiveness of the community. 
Table \ref{spatail} shows the average metrics values  of the communities detected for nodes whose communities detected by AppAcc are not empty.
Table \ref{spatail2} shows the the average metrics values  of the communities detected for all selected nodes.

Table \ref{spatail} shows that on most datasets, the AppSLDR algorithm has smaller $d_{avg}$ and $d_{IO}$ values than other methods, which means that nodes in the community found by AppSLDR have closer distances.
In terms of $d_{IO}$, AppAcc obtains the best results on the $Foursquare$ and $Syn2$ datasets.
For $d_{avg}$ and $d_{IO}$, the M method performs worse than the other comparison methods because it does not consider the spatial location information of nodes when detecting community structure.
Among the methods that consider the spatial location of nodes, Geomod finds communities with the largest values of $d_{avg}$ and $d_{IO}$ because Geomod detects all communities in the network from the perspective of global optimization. 
Table \ref{spatail2}  shows average values of metrics calculated for all selected nodes.
Although there are some differences between the values in Table \ref{spatail2} and those in Table \ref{spatail}, similar conclusions are drawn from Table \ref{spatail2}, i.e., the AppSLDR algorithm performs better than the M and Geomod methods.

As seen from Tables \ref{com}, \ref{com2}, \ref{spatail} and \ref{spatail2}, in terms of both structural and spatial cohesiveness, the AppSLDR algorithm outperforms the other comparison methods.

\subsection{Discussion}\label{sec:discussion}
\begin{figure}[!t]
	\centering
	\includegraphics[width=2.5in]{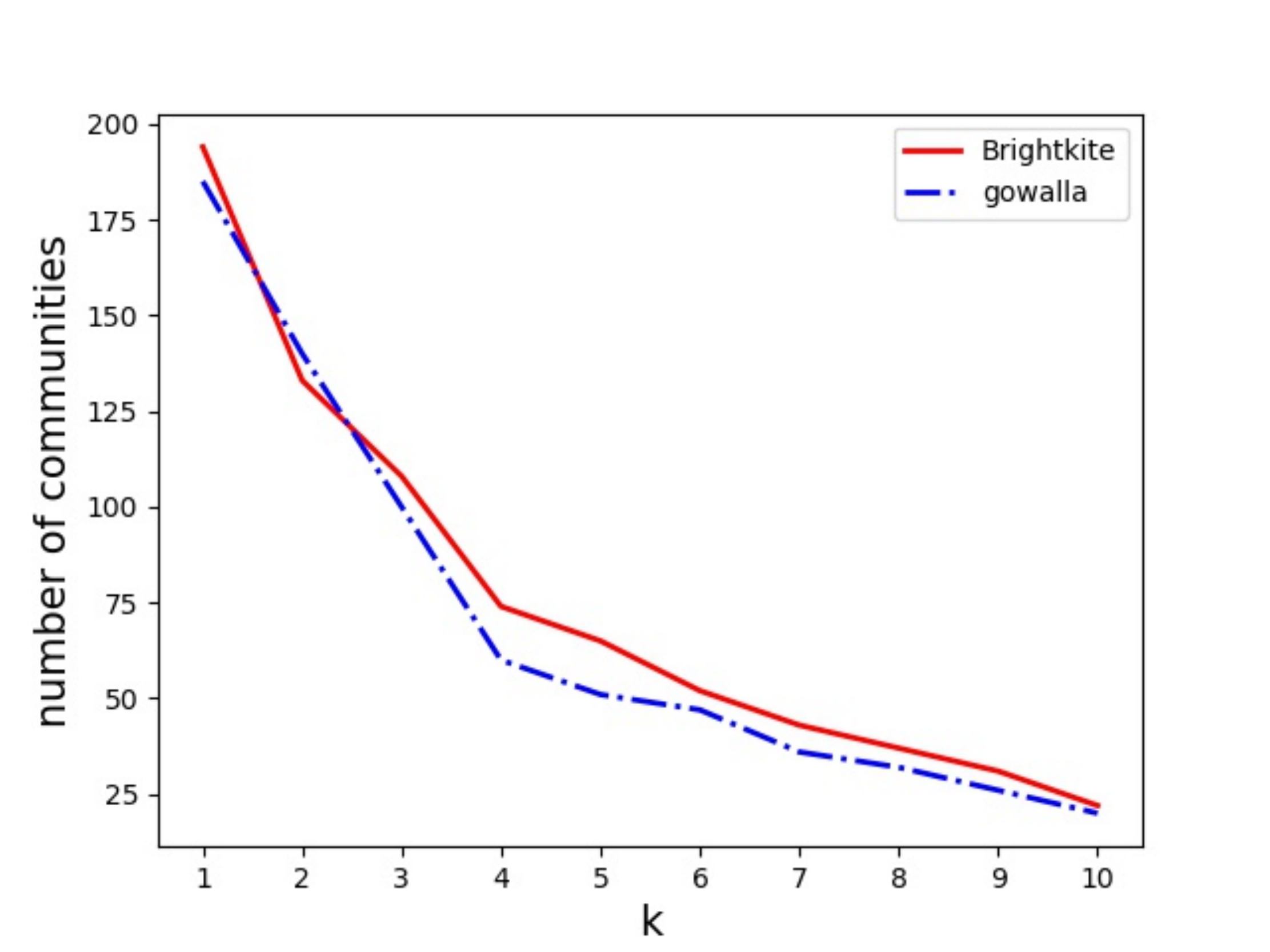}
	\caption{Number of communities detected by AppAcc as $k$ varies}
	\label{fig:number}
\end{figure}

As mentioned in Section \ref{sec:introduction}, the performance of community search algorithms is affected by the parameter $k$. We use the performance of AppAcc on the $Brightkite$ and $Gowalla$ datasets to
illustrate the effect of $k$ on the final results. A total of 200 given nodes are selected.
Fig.~ \ref{fig:number} shows the number of communities found by the AppAcc algorithm.
\subsubsection{Effect of $\ k$ on the AppAcc algorithm}
For the $Brightkite$ dataset, when $k$ = 1, 5 and 10, the number of communities is 200, 67 and 24, respectively. Correspondingly, the number of nodes for which no community is found are 0, 133 and 176. Similar results are obtained on the $Gowalla$ dataset.
From this phenomenon, we can see that $k$ greatly affects the performance of the AppAcc algorithm. Our algorithm has no parameters. For the 200 nodes selected, AppSLDR could find communities for each given node.
From this point of view, compared with AppAcc algorithm, AppSLDR is more robust.

Here, $expansion$ \cite{chakraborty2017metrics} is adopted to measure the sparsity of external edges of individual communities, calculated as follows:
\begin{equation}
	expansion=\frac{|E_C^{out}|}{|C|},
\end{equation}
where $|C|$ is the size of community $C$. The smaller the value of $expansion$ is, the better the structural cohesiveness of the community.
\subsubsection{Difference between AppAcc and AppSLDR in structural cohesiveness}
\begin{figure}[!t]
	\centering
	\includegraphics[width=2.5in]{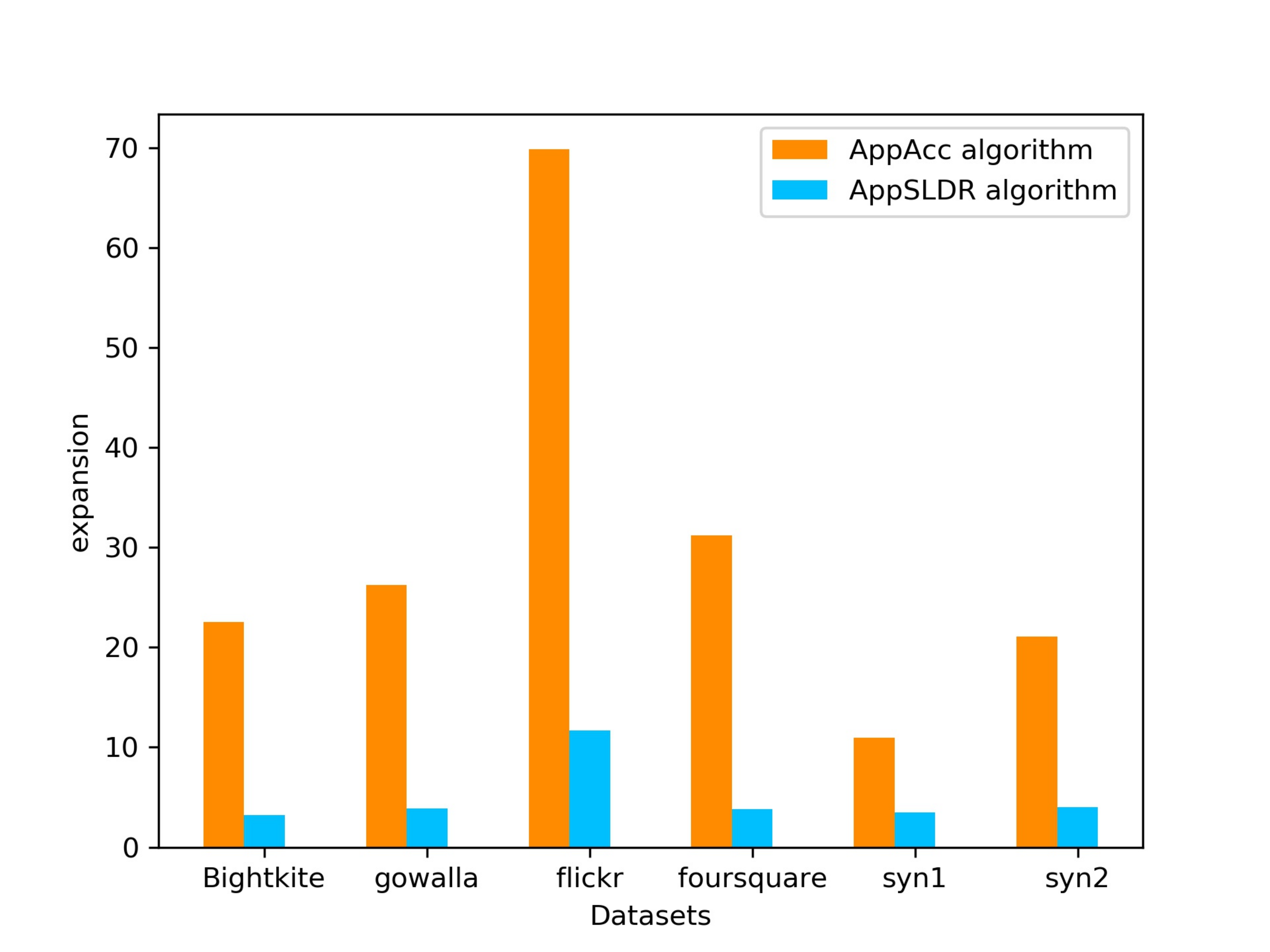}
	\caption{$expansion$ of AppAcc and AppSLDR}
	\label{fig:expansandsize}
\end{figure}
Fig.~\ref{fig:expansandsize} shows that the $expansion$ of AppAcc is tens of times larger than that of AppSLDR, which means that the community detected by AppSLDR is more sparsely connected to the external nodes.
The reason is that AppAcc considers only the closeness within the community, while AppSLDR considers the closeness within the community as well as the differences inside and outside the community.

\subsubsection{Comparison of SLDR and AppSLDR}
\begin{table*}[!t] 
	\centering
	\fontsize{10}{15}\selectfont
	\small
	\caption{Comparison of SLDR and AppSLDR}
	\renewcommand\arraystretch{1.1}
	\label{tab:base}
	\begin{threeparttable}\setlength{\tabcolsep}{4.6mm}{
    	\begin{tabular}{c|c|c|c|c|c|c}
    		\toprule
    		\multirow{2}{*}{\diagbox{Dataset}{Method}}&
    		\multicolumn{3}{c|}{SLDR}&\multicolumn{3}{c}{AppSLDR}\cr\cline{2-7}
    		&time(s)&$communitude$&$d_{avg}$&time(s)&$communitude$&$d_{avg}$\cr
    		\hline
    		$Syn1$	& \rmfamily 2121.0 & \rmfamily 0.278&\bfseries\rmfamily 0.205 & \bfseries\rmfamily 224.6& \bfseries\rmfamily 0.353 & \rmfamily 0.285 \cr
    		
    		$Brightkite$ & \rmfamily 3081.5 & \rmfamily 0.488 & \bfseries\rmfamily 0.007 &\bfseries\rmfamily 1408.4 & \bfseries\rmfamily 0.550&\rmfamily 0.012\cr
    			
    		\bottomrule
    	\end{tabular}}
	\end{threeparttable}
\end{table*}

We also compare AppSLDR with the SLDR algorithm in terms of runtime, $communitude$ and $d_{avg}$. Due to the slow speed of SLDR, we only compare SLDR and AppSLDR on two datasets, $Syn1$ and $Brightkite$. Table \ref{tab:base} shows the results of the SLDR and AppSLDR algorithms. We observe that $d_{avg}$ of the SLDR algorithm is slightly better than that of the AppSLDR algorithm. Although AppSLDR loses little spatial cohesiveness, it achieves almost several times faster speed. Moreover, the $communitude$ of the AppSLDR algorithm is better than that of the SLDR algorithm.

Specifically, we analyzed the runtime of the algorithms on the 200 nodes. On the $Brightkite$ ($Syn1$) dataset, for AppSLDR algorithm,
the runtime of 59.5\% (90\%) nodes is within 10 seconds, and the runtime of 14.5\% (3\%)  nodes is greater than two hours, which had a significant impact on the average results. 
% \textcolor{red}{Therefore, to reflect the actual runtime of most nodes, if a detection is running for more than two hours, its runtime will not be counted.}
As a comparison, for SLDR algorithm, only 36\% (21.5\%) nodes whose runtime is within 10 seconds on the $Brightkite$ ($Syn1$) dataset. 
The reason for the runtime of some nodes more than two hours is as follows: The degrees of these nodes or their neighbor nodes are greater than several thousand, which causes our algorithms to spend much overhead to generate derived communities.

\section{Conclusion}\label{sec:conclusion}
In this paper, we study the SLCD problem, which aims to detect a spatial-aware local community with only local information.
To address this problem, we propose the SLDR algorithm and its efficient approximation algorithm called AppSLDR.
Extensive experiments on synthetic and real-world datasets demonstrate that the AppSLDR algorithm substantially outperforms other methods in both structural and spatial cohesiveness.
In the future, we plan to extend the SLDR and AppSLDR algorithms to detect the community structure in attribute networks.

\bibliography{paper}

\end{document}